\documentclass[]{interact}

\usepackage{soul} %for adding correction / striketroughs
\usepackage{color}
\newcommand{\red}{}

\usepackage{epstopdf}% To incorporate .eps illustrations using PDFLaTeX, etc.
\usepackage{subfigure}% Support for small, `sub' figures and tables
\usepackage{multirow}

\usepackage[numbers,sort&compress,merge]{natbib}% Citation support using natbib.sty
\bibpunct[, ]{(}{)}{,}{n}{,}{,}% Citation support using natbib.sty
\theoremstyle{plain}% Theorem-like structures

\theoremstyle{definition}

\theoremstyle{remark}

\newcommand{\mum}[1]{$\mu$m#1}
\newcommand{\Ca}{$^{40}\mathrm{Ca}^+$~}
\newcommand{\figref}[1]{Fig.~\ref{#1}}

\graphicspath{{./Figures/}}

\let \mr=\mathrm

\usepackage{amsmath}	% required for `\align' (yatex added)
\begin{document}
	
	%\articletype{DRAFT}
	
	\title{Precise positioning of an ion in an integrated Paul trap-cavity system using radiofrequency signals}
	
	\author{
		\name{Ezra Kassa$^{\ast}$\thanks{$^{\ast}$Email: Ezra.Kassa@sussex.ac.uk}, Hiroki Takahashi, Costas Christoforou and Matthias Keller}
		\affil{Department of Physics and Astronomy, University of Sussex, Brighton, BN1 9RH, United Kingdom}
	}
	
	\maketitle
	
	\begin{abstract}
		We report a novel miniature Paul ion trap design with an integrated optical fibre cavity  which can serve as a building block for a fibre-linked quantum network.
		In such cavity quantum electrodynamic set-ups, the optimal coupling of the ions to the cavity mode is of vital importance and this is achieved by moving the ion relative to the cavity mode. The trap presented herein features an endcap-style design complemented with extra electrodes on which additional radiofrequency voltages are applied to fully control the pseudopotential minimum in three dimensions. This method lifts the need to use three-dimensional translation stages for moving the fibre cavity with respect to the ion and achieves high integrability, mechanical rigidity and scalability. 
		Not based on modifying the capacitive load of the trap, this method leads to precise control of the pseudopotential minimum allowing the ion to be moved with precisions limited only by the ion's position spread. We demonstrate this by coupling the ion to the fibre cavity and probing the cavity mode profile. 
		
	\end{abstract}
	

	\section{Introduction}

	The field of atomic physics has advanced greatly since the advent of ion traps which confine ions for unprecedented durations without utilising the internal states of the ions.
	Because ion traps offer unparalleled levels of control over the ions' mechanical and internal degrees of freedom, many experiments have sought to combine them with optical cavities for enhanced atom-light interactions. As a result, there have been a number of significant experiments: single photons were generated on demand \cite{Keller:04}, %single cubits have been coupled to a high-finesse cavity \cite{bibid},
	cavity sideband cooling was performed on single ions \cite{Leibrandt:09}, super-radiance was observed with the collective coupling of coulomb crystals \cite{Herskind:09}, tunable ion-photon entanglement has been demonstrated \cite{Stute:12},
	%the collection of photons by ion-cavity systems was studied \cite{Sterk:12},
	multiple ions have been deterministically coupled to a cavity \cite{Begley:16}.  
	%study enhanced atom-light interactions at the single quanta level by integrating them with optical cavities \cite{}. 
	The combination of ion traps with optical cavities is also considered to be one of the most promising avenues for advances in quantum information processing. 
	%Quantum computers are expected to perform certain tasks which no classical computers can do in practical time scales.
	%A demonstrative quantum computer requires in the excess of $10^5$ quantum bits (qubits) \cite{Martinis:15,Fowler:12}.
	Whilst there has been remarkable progress in the preparation, gate operation and readout of qubits \cite{Harty:14,Harty:16}, to date, these implementations have been limited to small scales, with 14 being the largest number of qubits entangled \cite{Monz:11}. 
	%Decoherence playing a crucial role in 
	Presently, challenges in the physical implementations of large quantum systems pose the greatest difficulty in advancing experimental quantum information science. 
	%The challenge being decoherence, low entanglement distribution efficiency, 
	Among the proposed solutions to tackle the scalability problem (eg.\cite{Lekitsch:17, Crick:10, You:02, Cirac:97}), distributed quantum information processing based on photonic links is the most promising.
	%Challenges in the physical implementations of large quantum systems has limited the largest number of qubits entangled to date at 14 \cite{Blatt group}. 
	Notably, modular approaches using trapped ions as stationary qubits have attracted significant interest. %due to their unparalleled quantum control within small-scale ion traps\cite{Hucul:15,Nigmatullin:16}.
	However, so far, optically heralded entanglement with remote trapped ions has only been demonstrated using high numerical aperture lenses for the collection of photons, a method which suffers from low efficiencies in the entanglement generation\cite{Hucul:15,Moehring:07}. Placing the ions in an optical cavity, this efficiency can be greatly enhanced. 
	%IN THIS SCENARIO, THE POSITIONING OF THE ION WITH RESPECT TO THE CAVITY MODE IS OF PARAMOUNT IMPORTANCE.
	Further, by reducing the cavity mode volume, one can enhance the ion-cavity coupling and, subsequently, the efficiency of operations. To this end, fiber-based Fabry-P\'erot cavities (FFPCs) have been combined with ion traps \cite{Steiner:13,Ballance:17,Brandstatter:13}.
	In such ion trap-cavity systems, the optimal positioning of the ion with respect to the cavity mode is of vital importance.  
	In the previously demonstrated designs of ion traps combining FFPCs \cite{Brandstatter:13,Steiner:13,Ballance:17}, the FFPCs were mechanically translated to optimise the overlap between the ion and the cavity mode.  
	In addition to the need for a three-dimensional positioning system which tends to be bulky and expensive, the movable cavities affect the trapping field and shift the geometrical center of the trap as they are moved. This adds complexity to the trapping and optimisation of the ion-cavity coupling. 

	In contrast, in this paper, the tuning of the ion-cavity coupling is done by moving the ion electrically instead of moving the cavity.

	In this way, we minimise the
	number of the movable elements in our trap and realise a highly integrated system
	which facilitates mechanical rigidity, compactness and scalability without sacrificing
	the ability to reach optimal ion-cavity coupling.
	Recently this trap was successfully used to demonstrate the alteration of photon emission rates of a single trapped ion due to the Purcell effect \cite{Takahashi:17}.  
	
	Moving an ion in a Paul trap using a dc voltage is often not acceptable as it causes heating of the ion through excess micromotion. An approach based on the modification of the pseudopotential has been established in \cite{Herskind:09_2} by using tunable capacitive loads attached to the trap electrodes. The authors report a micrometer precision in their ability to shift the ion's position.  
	However the drawbacks in this method are: 1) The secular frequency of the trap may be substantially lowered by the added capacitive load. 2) Care  must be taken to minimise the phase shift on the rf signal caused by the capacitors which would otherwise lead to excess micromotion. 3) Manual tuning of bulk capacitors is difficult to repeat with precision.  
	%Radiofrequency (rf) in Paul traps is generally delivered by a resonant circuit and a drawback of adding a parallel load is that the resonance frequency of the trap may be substantially lowered. This is undesirable in practice since it results in the lowering the secular frequency as well as the trap depth. One can add a series capacitive load to reduce the change in resonance frequency, however careful adjustments must be made to minimise the phase shift between the trapping electrodes which would otherwise lead to excess micromotion.  
	In contrast to the techniques employed in \cite{Herskind:09_2}, our trap architecture presented here does not require the addition of capacitors to shift the pseudopotential minimum. Instead, additional rf signals are directly applied to rf-electrodes integrated within the trap. The amplitudes and phases of the additional rf signals are digitally controlled by function generators, allowing for rapid, precise and repeatable tuning. As a result, we have been able to improve the positioning of the pseudopotential minimum down to the spread of the ion position.  
	
	This paper is organised as follows. In section 2, we introduce the core trap design featuring extra electrodes used to move the ion electrically. We numerically simulate the ion's motion with respect to the cavity mode centre. In section 3, we describe the trap infrastructure with a focus on the integrability and mechanical rigidity of the design which is attested by the measured stability of the FFPC. In section 4, we present the experimental methods used to compensate excess micromotion due to the additional rf fields. We show the mapping of the cavity mode by displacing the ion. We conclude with a summary and outlook in section 5.

	%IN DREWSEN PAPER, CHANGE OF  ION POSITION IS DISCRETELY DONE BY MEANS OF PHYSICALLY CHANGING THE CAPACITIVE LOADS USED.
	%ONE CAN ADD A SERIES LOAD TO KEEP THE CHANGE IN RESONANCE FREQUENCY SMALL, HOWEVER CAREFUL ADJUSTMENTS MUST BE MADE TO MINIMISE THE PHASE SHIFT BETWEEN ELECTRODES LEST EXCESS MICROMOTION INCREASES.  

	\section{Ion trap design}
	
	\begin{figure}[h]
		\centering
		\includegraphics[width=1\linewidth]{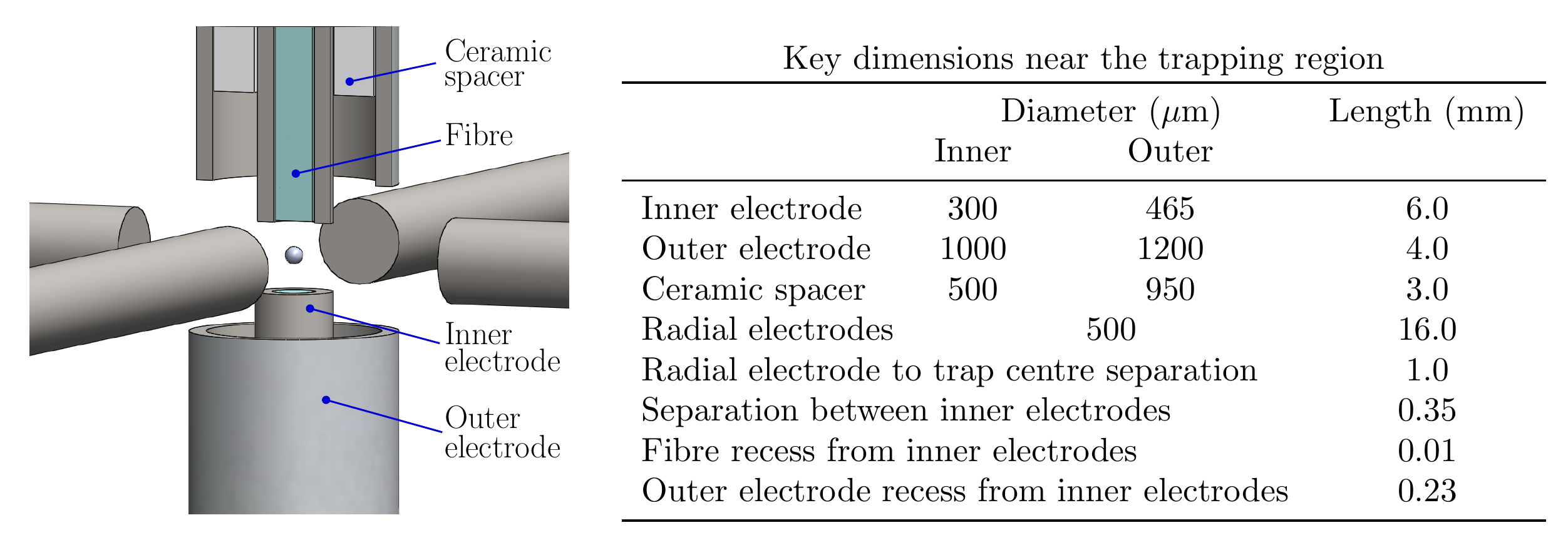}
		\caption{A drawing of the trapping region. A pair of electrode assemblies face one another in the axial direction, surrounded by four electrodes on the radial plane. Only a cross section of the upper assembly is shown to reveal the inner structure. The ion is symbolised by a small sphere in the centre. The main pseudopotential is generated by applying an rf voltage to the outer electrodes whilst the other electrodes are held at rf-ground.}
		\label{fig:trap_cutout}
	\end{figure}
	
	We have designed a novel radio-frequency (rf) Paul trap with an integrated FFPC. The trapping structure is an endcap-style ion trap \cite{Takahashi:13,Kassa:17}. The trapping potential is formed by a pair of electrode assemblies each consisting of two concentric stainless steel tubes separated by a ceramic spacer (see \figref{fig:trap_cutout}). The assembly is bonded together by an ultra-high vacuum compatible adhesive. The key design dimensions of components near the trapping region are detailed alongside the schematic in \figref{fig:trap_cutout}.
	%The outer electrode of both assemblies are recessed by 230 \mum  with respect to the inner electrode. The inner electrodes of the opposing assemblies are separated by 350 \mum.
	By applying an rf voltage on the outer electrodes whilst holding the inner electrodes at rf-ground, a trapping potential can be formed to trap a single ion at the centre of the design. The cavity fibres are inserted into the inner electrodes such that the resulting cavity mode encompasses the trapped ion. By recessing the fibres by ~5-10 \mum{}  in the inner electrodes, the dielectrics are well shielded from the trapping electric field and do not distort the trapping potential as a result.

	On the radial plane are four cylindrical electrodes placed 1.0 mm from the trap centre. By applying dc voltages to two adjacent electrodes as well as the inner electrodes, the excess micromotion due to stray static electric charges can be fully compensated in three dimensions (3D) using the rf-correlation technique \cite{Berkeland:98}. By applying rf signals synchronous to the main trap drive to the other two radial electrodes as well as the inner electrode\red{s}, the pseudopotential minimum position can be fully controlled in 3D\red{,} allowing the ion-cavity coupling to be optimised both radially and axially.
	
\begin{figure}[h]
		\centering
		%\subfigure[]{		%{\includegraphics[width=0.47\linewidth]{MainPotential_Slice}}}
		\subfigure[]{
			{\includegraphics[width=0.47\linewidth]{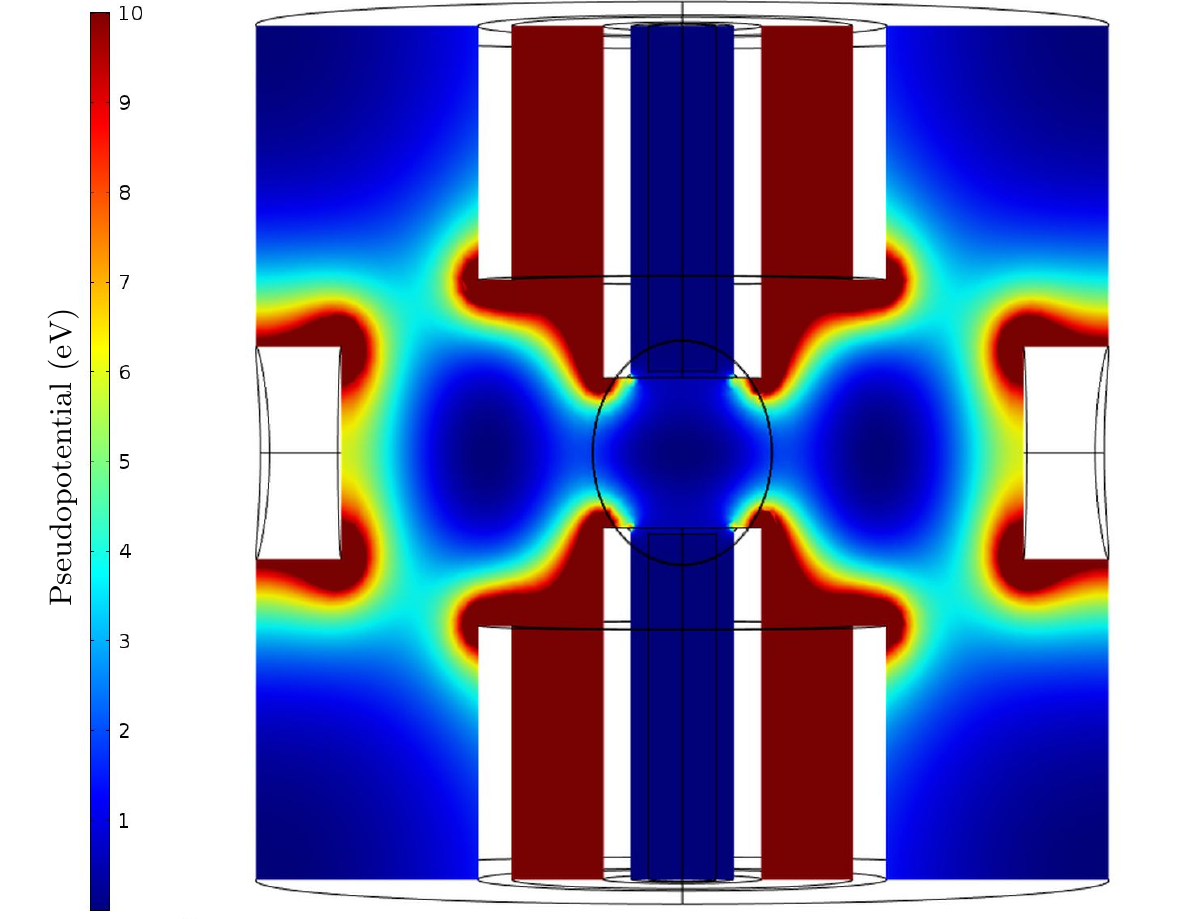}}}
		%\subfigure[]{
		%{\includegraphics[width=0.47\linewidth]{MainPseudopotential_CrossSections}}}
		\subfigure[]{
			{\includegraphics[width=0.475\linewidth]{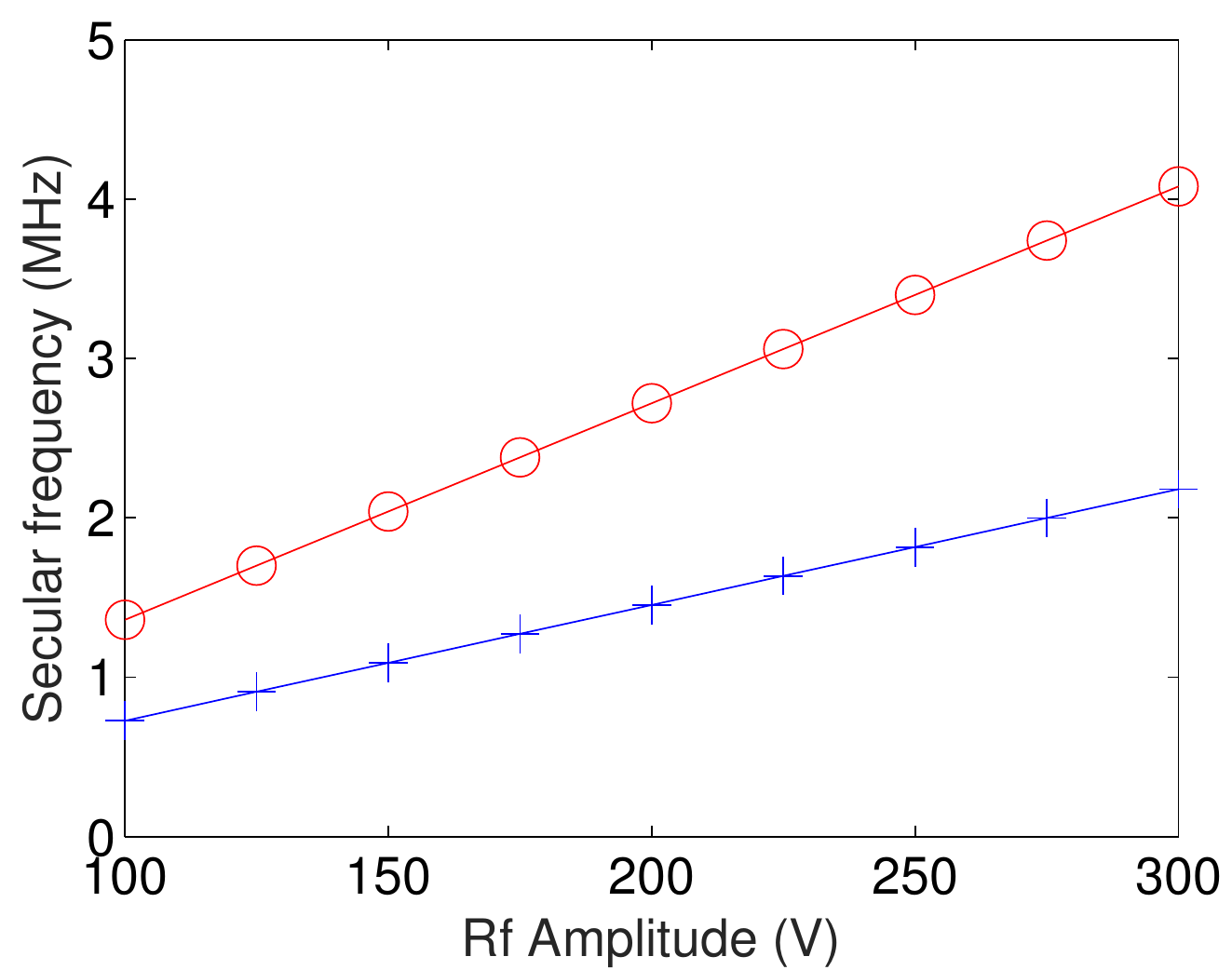}}}
		\caption{%(a) The electrostatic potential when 200 V is applied to the outer electrodes. All other electrodes are held at ground.
			(a) The resulting pseudopotential when the applied field is alternating at 20MHz. The plotted pseudopotential is capped at 10 eV to emphasise its structure near the trapping region.
			%(c) Cross-sectional plots of the pseudopotentials in (b). The kinks in the lines (eg. at $\pm0.3$mm) are artefacts of poor mesh-refining in the simulations. The sharp drop in the axial pseudopotential at $\pm0.19$mm is due to meeting the fibre boundary.
			(b) Secular frequencies for different drive amplitudes are simulated for the radial (red circles) and axial (blue crosses) directions and fitted (lines).} \label{fig:Main_Potential_and_Pseudopotential}
	\end{figure}

	%The simulations \cite{Comsol} in
	\figref{fig:Main_Potential_and_Pseudopotential} shows a simulation \cite{Comsol} of  the trap pseudopotential when an rf voltage at a frequency of 20 MHz and amplitude 200 V is applied.
	%to the outer electrodes while the inner electrodes are held at rf-ground.
	In this simulation, all radial electrodes are held at rf-ground. One can see that the fibres are well-shielded from electric fields. The pseudopotential at the trap centre can be well approximated by the quadratic potential of an ideal Paul trap. Although, the radial electrodes break the cylindrical symmetry, they have negligible effect near the centre. %as can be seen in the cross-sectional plots shown in \figref{fig:Main_Potential_and_Pseudopotential}(c).
	From fits to the pseudopotentials at different rf voltage amplitudes (see \figref{fig:Main_Potential_and_Pseudopotential}(b)), we find axial and radial secular frequencies of 7.3 kHz/V and 13.6 kHz/V respectively.
	
\begin{figure}[h]
		\centering
		\subfigure[]{
			{\includegraphics[width=0.45\linewidth]{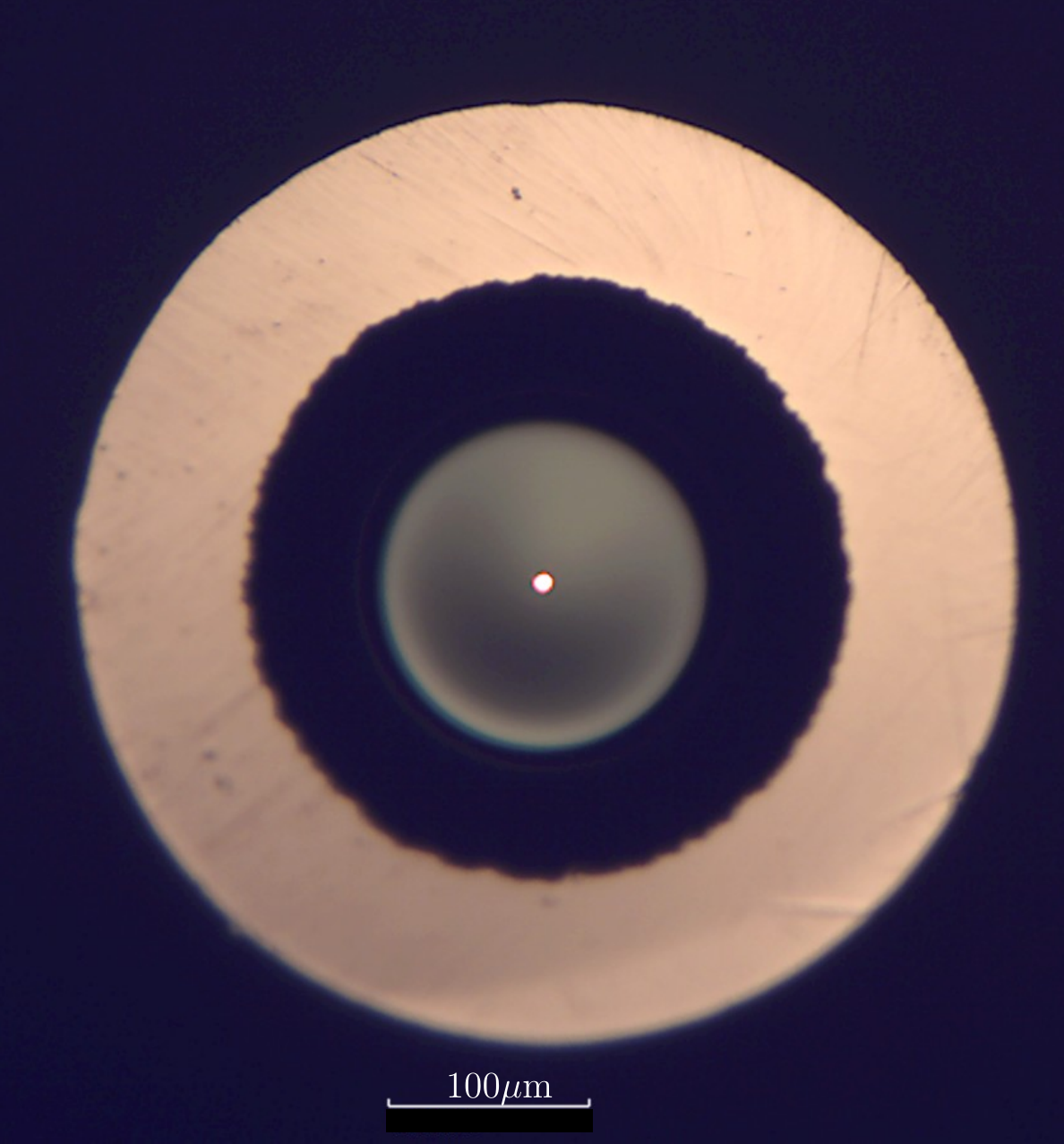}}}
		\subfigure[]{
			{\includegraphics[width=0.45\linewidth]{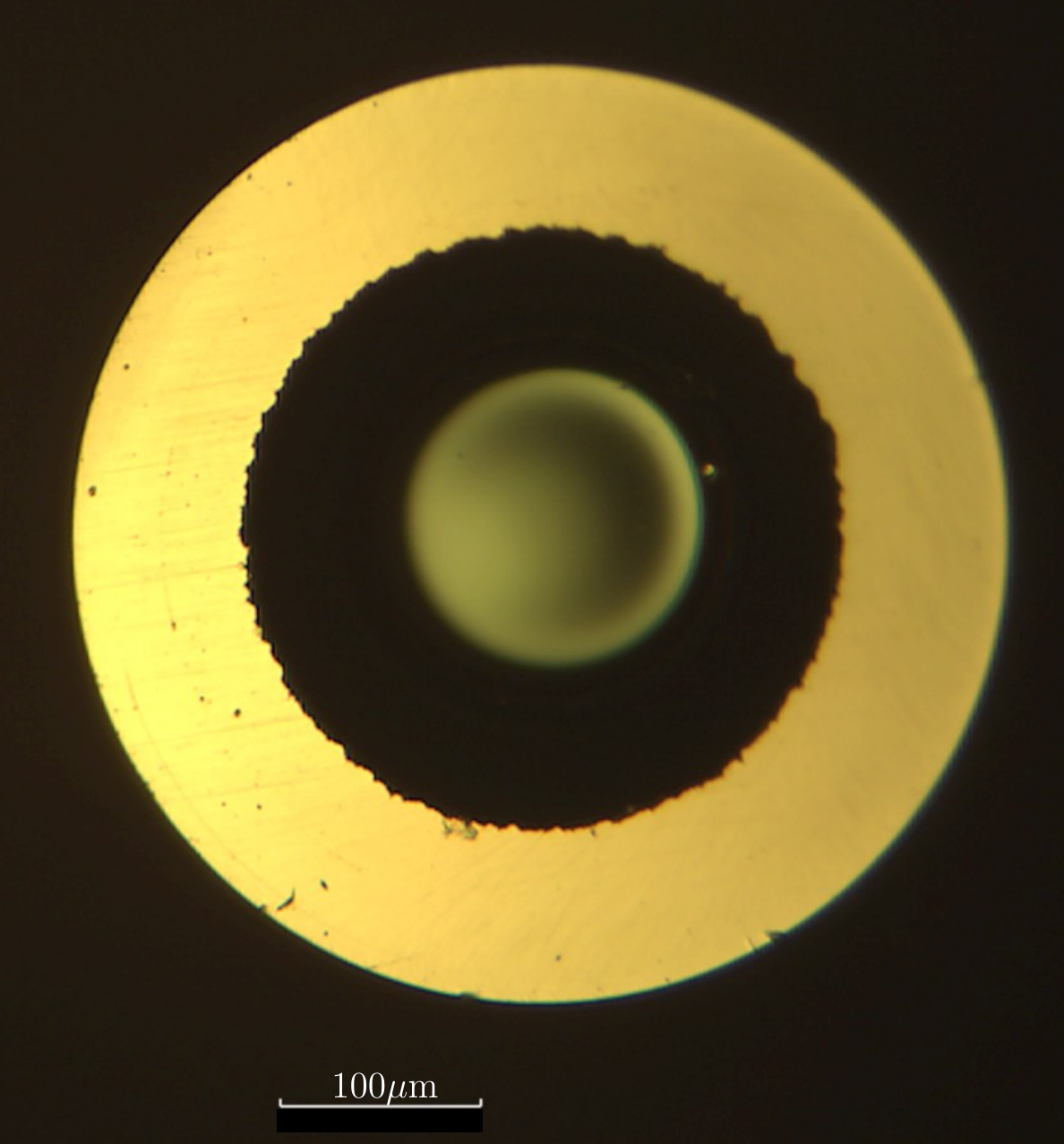}}}
		\caption{(a) A facet image of the SM fibre inside the upper inner electrode. The core of the fibre is illuminated by a laser beam injected into the fibre. (b) The MM fibre inside the inner electrode of the lower assembly.} \label{fig:FibreConcentricity}
	\end{figure}

	The FFPC is formed by two fibres \cite{Takahashi:14} whose facets have radii of curvature of 560 \mum{} and are coated for high reflectivity at 866 nm wavelength with 25 ppm transmission resulting in a finesse of 48,000. At the chosen cavity length of 370 \mum{}, the cavity mode waist is calculated to be 8.5 \mum{}.  The fibre in the upper assembly is a single mode (SM) fibre and serves as the input. The second fibre is a multimode (MM) fibre and serves as the output %The SM fibre has a cladding diameter of 200 \mum{} and the MM one, 198 \mum{}.
	\red{as the cavity mode is most efficiently coupled into this fibre}.
	The SM and MM fibres have a protective copper coating with 
	%diameters 260$\pm10$ \mum{} and 240$\pm10$ \mum{}
	thicknesses of 30 \mum{} and 20 \mum{} respectively. The FFPC fibres are carefully inserted and glued to the electrode assemblies ensuring the pair are concentric. However, due to imperfections in the insertion procedure and due to thermal drifts during the curing period, the axes of symmetry of the inner electrode and the fibre may deviate from one another by a few micrometers. Facet images of the inner electrodes with the inserted fibres (see \figref{fig:FibreConcentricity}) reveal off-concentricities of 3 $\pm$ 1 \mum{} and 9 $\pm$ 1 \mum{} for the assemblies. As a result, we cannot expect the trapped ion to optimally overlap with the cavity mode therefore highlighting the need for the radial electrodes on which rf voltages can be applied to shift the ion's radial position.
	%
	%
	%We simulate the pseudopotential when an rf voltage is applied to one of the radial electrodes in addition to outer electrodes.
	
\begin{figure}[h!]
		\centering
		\subfigure[]{
			{\includegraphics[width=0.45\linewidth]{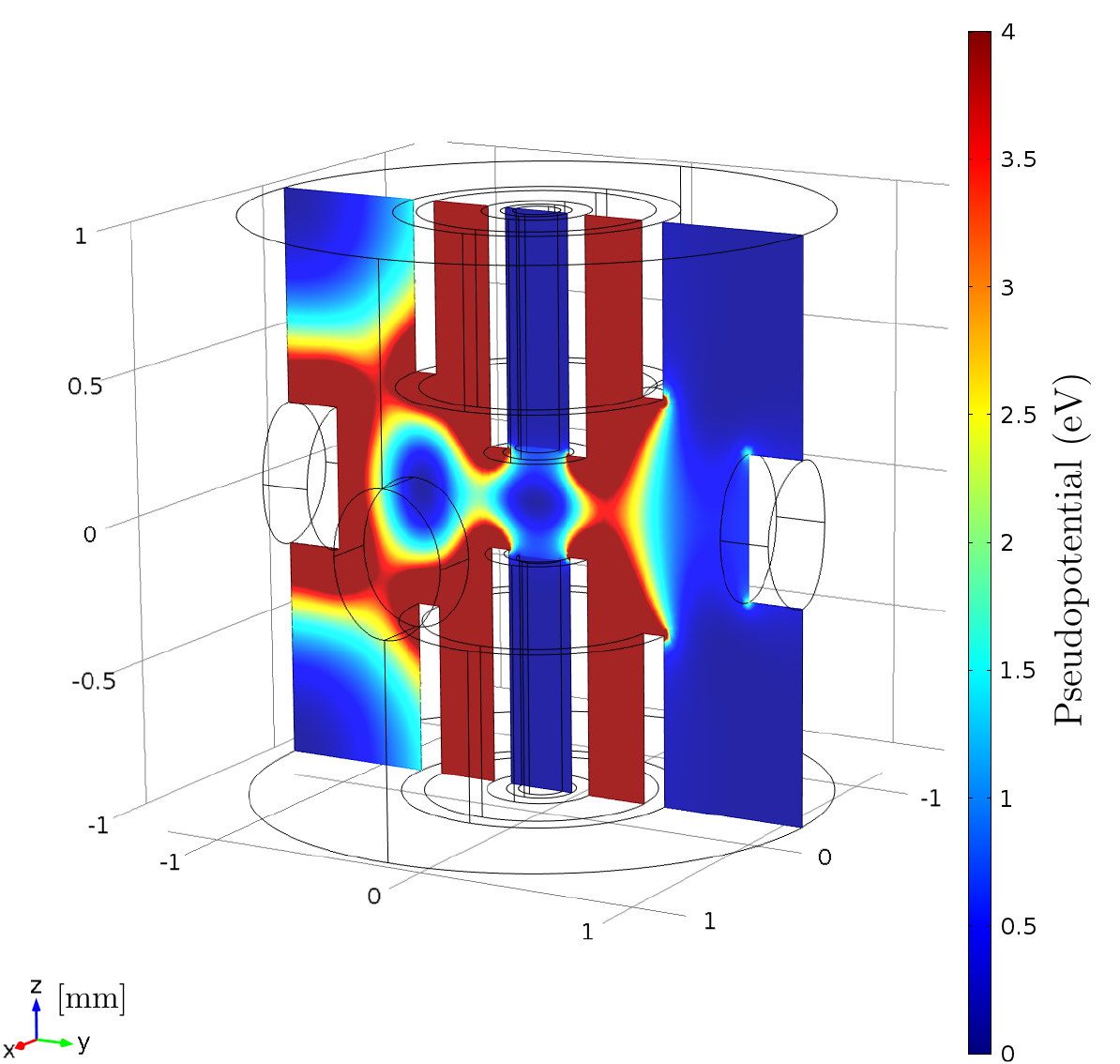}}}
		\subfigure[]{
			{\includegraphics[width=0.5\linewidth]{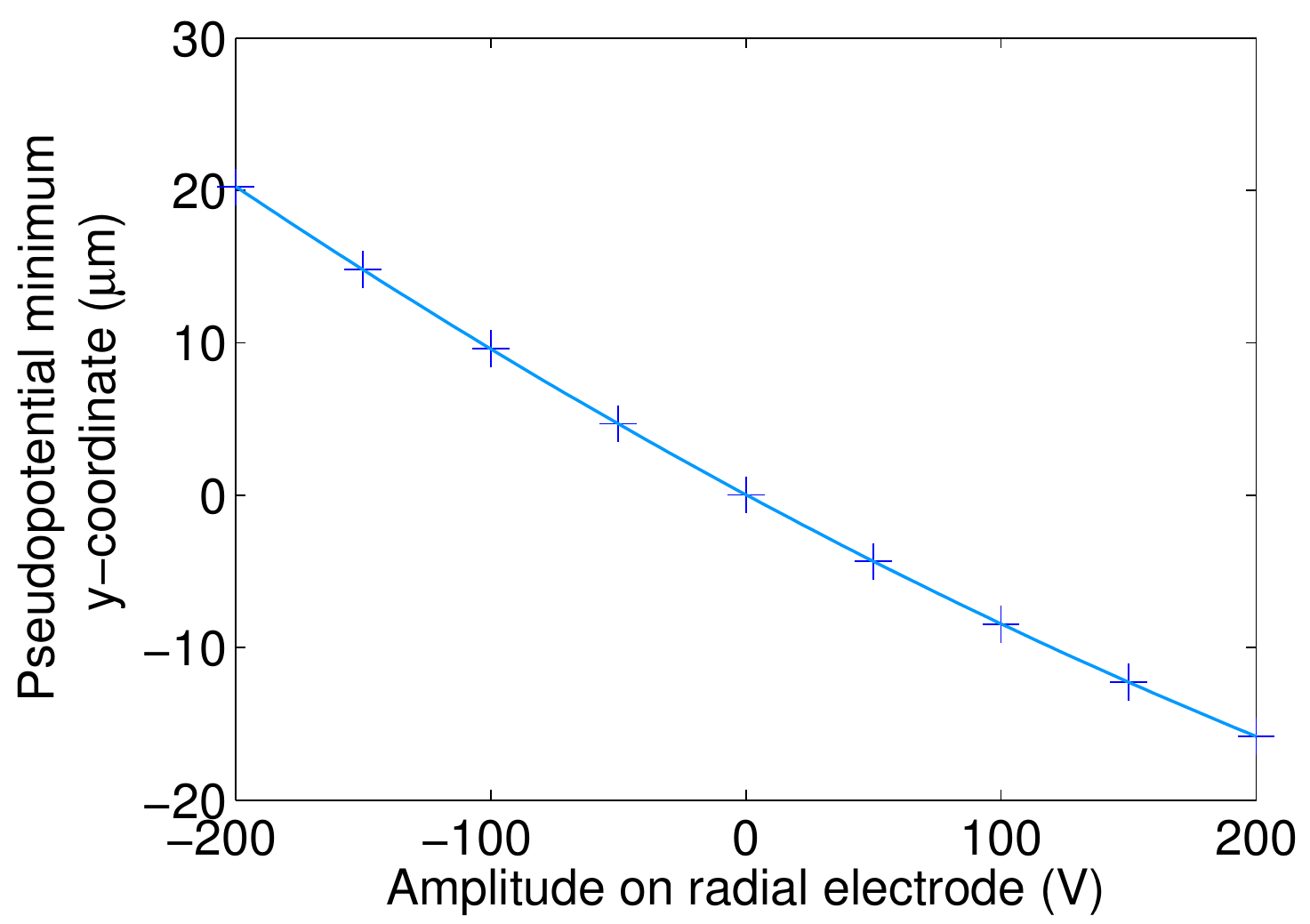}}}
		\caption{(a) The pseudopotential when an additional rf voltage of amplitude 200V synchronous to the main drive is applied to the radial electrode in the positive y-direction.  (b) The pseudopotential minimum position versus the rf-amplitude applied on the radial electrode. In dark blue crosses are the simulated data points. The light blue line is a second order polynomial fit.} \label{fig:Pseudopotential_min_vs_sideRF}
	\end{figure}

	\figref{fig:Pseudopotential_min_vs_sideRF}(a) shows a cross section of the pseudopotential when an rf voltage of 200 V is applied to one of the radial electrodes. At a typical main drive amplitude of 200 V, we find that the ion can be shifted radially by more than 15 \mum{}. %with the same voltage amplitude on one of the radial electrodes.
	From the pseudopotential minimum positions, $y_{min}$, at various rf amplitudes, $V_{y}^{\mathrm{rf}}$, on the radial electrode (see \figref{fig:Pseudopotential_min_vs_sideRF}(b)), we find the pseudopotential minimum position can be fitted with the second order polynomial $y_{min} =(6.1\times10^{-5} \mu\mr{m}/V^2) {V_{y}^{\mathrm{rf}}}^2 - (0.1\mu\mr{m}/V)V_{y}^{\mathrm{rf}}$.

	To precisely control the ion's position in the standing wave of the cavity mode, a \red{differential} rf voltage, $V_{z_u}^\mathrm{rf}$, can be applied on the inner electrode\red{s in an anti-symmetric manner (i.e.} $\red{V_{z}^{\mathrm{rf}}/2}$ \red{on the upper electrode and} $\red{-V_{z}^{\mathrm{rf}}/2}$ \red{on the lower electrode)}. With simulations we find that we can move the ion by more than 2 \mum{} (corresponding to more than two full wavelengths in the standing wave of the cavity more) with as little as 1 V of synchronous \red{differential} rf on the inner electrode\red{s}.

	\section{Trap infrastructure}

	\begin{figure}[h]
		\centering
		{\includegraphics[width=0.85\linewidth]{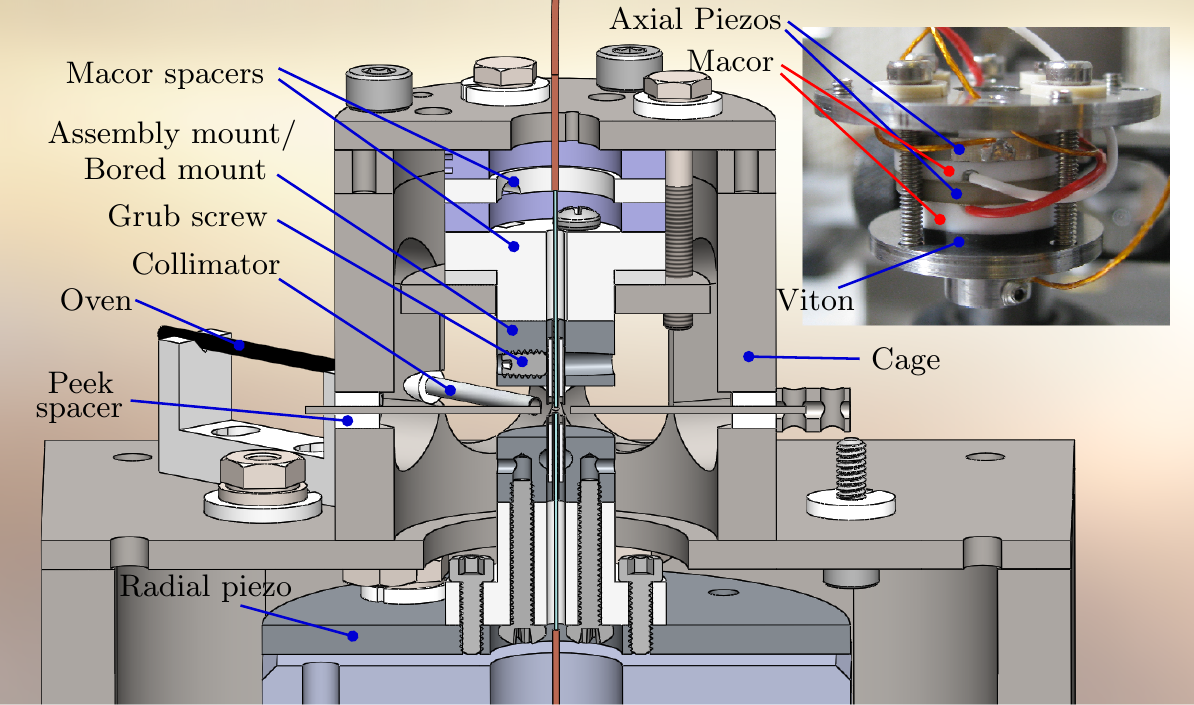}}
		\caption{A cross-sectional view of the surrounding architecture. An atomic oven is placed outside the \emph{cage} for thermal isolation. When resistively heated, an effusive beam is collimated by an electrode directing the atomic beam to pass between the electrode assemblies (where it gets ionised by lasers). Inset: A photograph of the upper assembly used in the experiment showing the stack of piezos. The upper(lower) piezo is the multilayer(monolayer) piezo. See main body for further discussion of the design. } \label{fig:Supporting_Structure}
	\end{figure}
	
	\figref{fig:Supporting_Structure} shows a cross section of the design supporting the trap electrodes. The main rf signal is supplied to the outer electrodes via bored cylindrical stainless-steel mounts to which the electrode assemblies are secured using grub screws. The lower assembly is attached to a piezoelectric translational stage \cite{PXY100} for precise radial alignment with respect to the upper assembly. %which itself is radially immobile.
	The radial alignment is done such that the cavity transmission signal is optimised.
	Electrical isolation for both electrode assemblies is provided by Macor platforms inserted between the bored stainless steel mounts and the rest of the supporting architecture. The inner electrodes are accessible through the Macor platforms and the bored openings for delivering electrical signals to them.
	
	Atop the Macor platform suspending the upper assembly are a pair of ring piezoelectric actuators  %(Noliac Multilayer NAC2124-C04, Noliac Monolayer NCE51-Ring-OD15-I9-TH2)
	%(Discuss piezos in more detail? Stroke and range? YES/NO)
	used to scan and stabilise the cavity length. One is a multilayer piezo \cite{Multi} with 
	%a range of 3.7 \mum{} and 
	stroke 1.5 nm/V whilst the other is a monolayer piezo \cite{Mono} with a 
	% range of 1.5 \mum{} and
	stoke of 0.45 nm/V. The combination allows for versatile feedback to the cavity. The scanning and stabilisation of the cavity length is facilitated by a Viton ring (VACOM) (see inset of \figref{fig:Supporting_Structure}) which compresses in response to change in pressure from the piezos. An additional Macor ring provides electrical isolation between the piezos. To aid the locking of cavity length, the system is isolated from surrounding mechanical vibrations by placing it on a set of stainless steel stacks separated by Viton rubber parts which inhibit the propagation of mechanical vibrations (not shown in the figure).
	Furthermore, the fibres are clamped immediately outside the structure to provide strain relief.
	
	A laser at 897 nm is used for stabilising the cavity length using the Pound-Drever Hall technique. The cavity linewidth with this laser is 22 MHz. For a stable cavity lock, the system should ideally have a flat response in gain and in phase for different feedback frequencies. This is however not the case in practice as the electrical as well as mechanical responses of system constituents (piezos, amplifiers, etc...) cause non-uniformity in the system's transfer function. As such, we first measured the transfer function of the system and found it to exhibit a derivative-like profile in gain accompanied by a sharp drop in phase at two distinct frequencies: 900 Hz and 9 kHz. %Such transfer function is a typical response of piezo-electric transducers mechanically coupled to other devices.
	This non-uniformity can be compensated by combining a low-pass filter, a band-pass filter and a high-pass filter. After introducing this custom filter, we are able to stably lock the cavity to the 897 nm laser to within a standard deviation of one thirteenth of the cavity linewidth. We find it is sufficient to use the multilayer piezo alone to stabilise the cavity.

	To trap ions, the oven is resistively heated expelling an effusive atomic beam  which is collimated to propagate between the electrode assemblies. The atomic beam is ionised by ionisation lasers at the trap centre in the trapping region which is optically accessible via the eight windows of the cage. 

	This design has been demonstrated to trap ions for several hours in the presence of the FFPC. 
	The cavity finesse of has not degraded over the course of 2 years.
	
	\section{Experiment}
	
	%\subsection{Cavity Locking}
	%
	\begin{figure}[h]
		\centering
		\subfigure[]{
			{\includegraphics[width=0.45\linewidth]{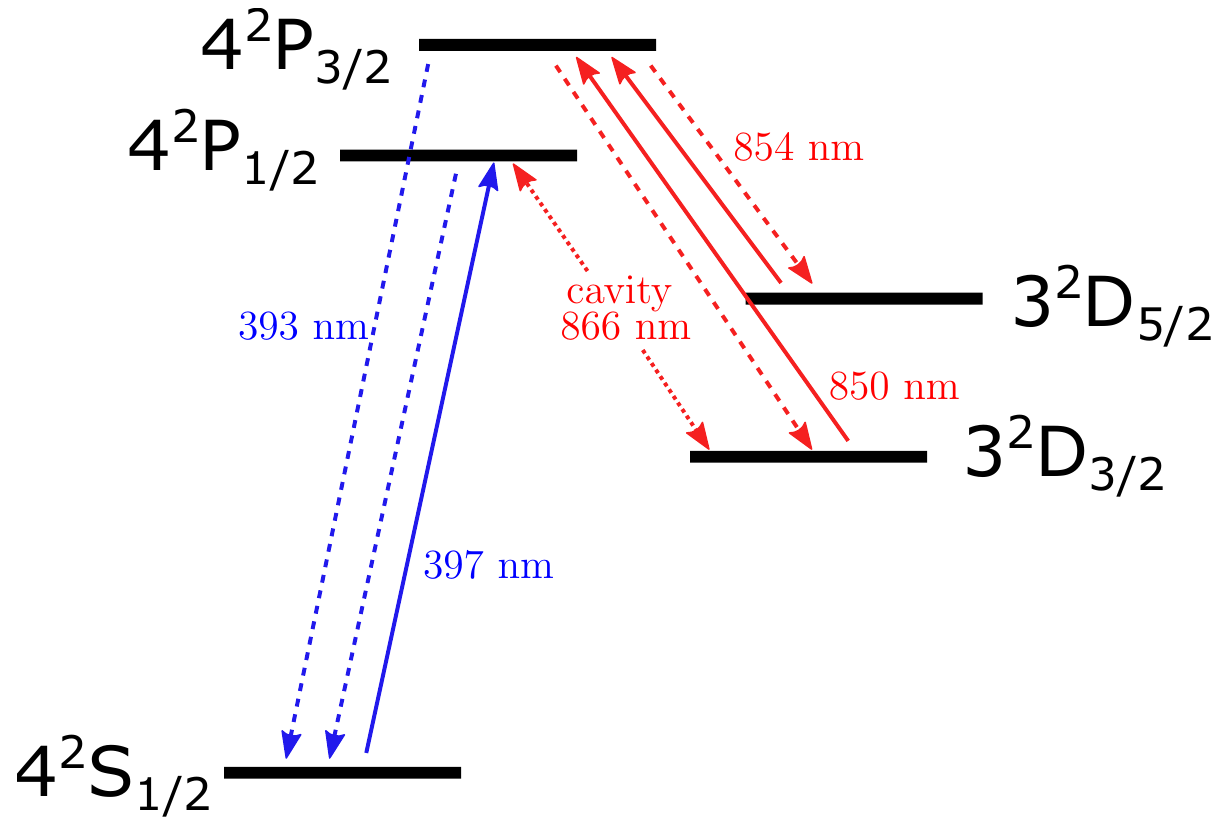}}}
		\subfigure[]{
			{\includegraphics[width=0.45\linewidth]{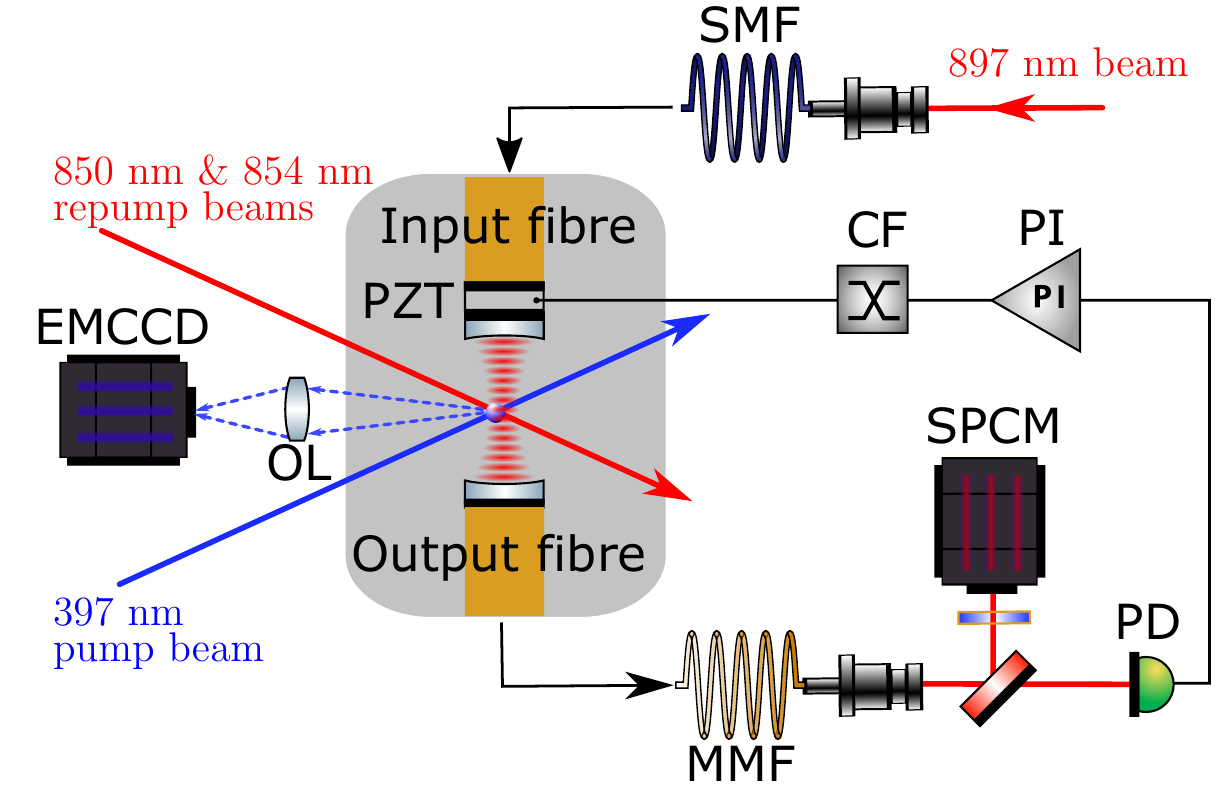}}}
		\caption{(a)The energy level scheme of \Ca. The wavelengths are given alongside relevant transitions marked. The solid arrows show transitions where lasers are employed. The dashed arrows show the spontaneous decay transitions. The cavity is tuned to the $P_{1/2} \leftrightarrow D_{3/2}$ transition (dotted arrows). (b) A simplified schematic of the optical set up. An objective lens (OL) focuses the ion's fluorescence onto an Electron Multiplying CCD camera (EMCCD) used to monitor the ion's position. A 897 nm laser is injected into the input fibre cavity via the singe mode fibre (SMF). The multimode fibre (MMF) collects the cavity output. A dichroic mirror is used to filter the 866 nm photons from the stabilising 897 nm beam which is transmitted to a photodetector (PD). To stabilise the cavity length, the signal from the PD is fed back to the piezos (PZT) via a proportional and integral (PI) circuit and a custom filter (CM). A single photon counting module (SPCM) is employed to count the cavity's 866 nm photon emission.}
		\label{fig:calcium40_and_optics}
	\end{figure}
	
	\subsection{Set up}
	
	The ionic species used in this experiment is \Ca. \figref{fig:calcium40_and_optics}(a) shows the relevant energy level scheme of the ion. The ion is Doppler cooled on the $S_{1/2} \leftrightarrow P_{1/2}$ transition. Repumper beams at 850 nm and 854 nm beams are used to depopulate the metastable $D_{3/2}$ and $D_{5/2}$ states respectively. Part of the ion's fluorescence is collected by a photomultiplier tube, the count rate of which is used to compensate excess micromotion using the rf-correlation technique as well as to perform spectroscopic measurements.   
	%The ion's position is monitored by an Electron Multiplying CCD camera (EMCCD) which co
	The ion's fluorescence is also focused onto an Electron Multiplying CCD camera (EMCCD) by a set of objective lenses (OL) (\figref{fig:calcium40_and_optics}(b)). The EMCCD image is used to monitor the ion's position.
	
	The fibre cavity is tuned to the $P_{1/2} \leftrightarrow D_{3/2}$ transition. The detection set up for the cavity emission is shown in \figref{fig:calcium40_and_optics}(b). An 897 nm beam is injected at the input of the single mode cavity fibre; the cavity transmission is used to stabilise the cavity length using the Pound-Drever Hall technique. The frequency of the 897 nm beam is finely tuned such that the cavity satisfies a double resonance condition for both the stabilising beam and the $P_{1/2} \leftrightarrow D_{3/2}$ transition at 866 nm.
	To separate the cavity emission signal at 866nm from the stabilising beam (897 nm) a dichroic mirror and band pass optical filters are employed. The cavity emission at 866 nm is detected by a single photon counting module (SPCM) whilst the stabilising beam is transmitted toward a photo-detector (PD) whose signal is fed to the piezo-electric transducers (PZT) via feedback electronics to stabilise the cavity length. 
	
\begin{figure}[h]
		%\centering
		\hspace{1cm}
		\includegraphics[width=0.8\linewidth]{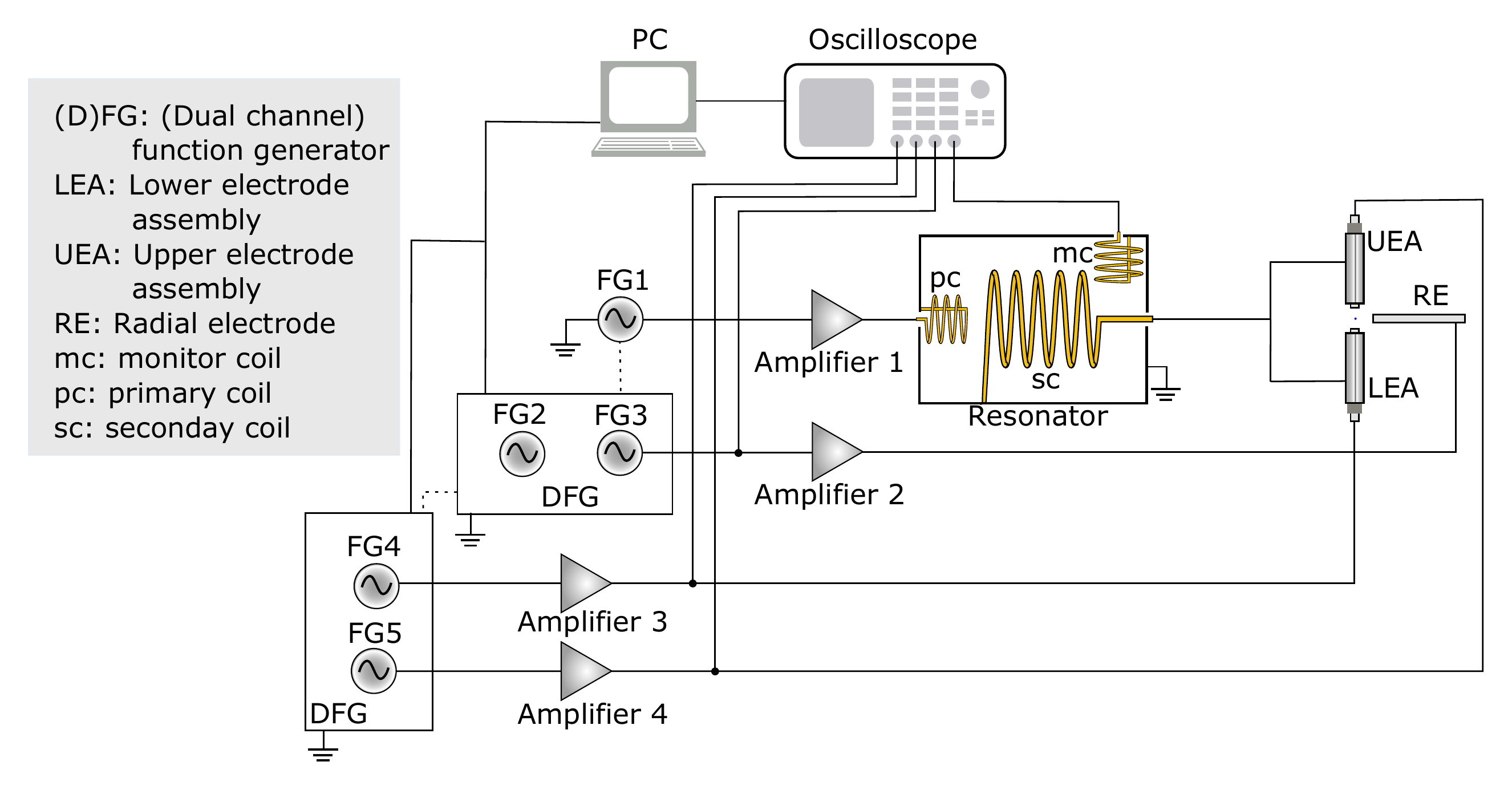}
		\caption{A simplified schematic of the electrical set up. Function Generator 1 (FG1) is used to drive the main electrodes via an amplifier (Amplifier 1) and a helical resonator. The phase at the output of the resonator can be gauged using the monitor coil (mc). The two channels of a dual function generator (DFG)  \red{are} used to apply additional rf voltages to the two rf radial electrodes. Only one radial electrode (RE) \red{is shown} for simplicity. Another \red{dual} function generator (FG4) is used to apply \red{a differential} rf signal to the inner electrode\red{s} of the  electrode \red{assemblies}. An oscilloscope monitors the phase of the various signals. FG2, FG3, \red{FG4 and FG5} are clock-synchronised to FG1. See the main text for further description. }
		\label{fig:MovingIonWithRFSetup_JMO}
	\end{figure}

	The schematic in \figref{fig:MovingIonWithRFSetup_JMO} shows the electrical set-up of the system. For simplicity, we only show one radial electrode and the dc connections are not included.  A helical resonator is used to amplify the main rf drive delivered to the outer electrodes of the upper electrode assembly (UEA) and the lower electrode assembly (LEA). Function generators FG2, FG3, \red{FG4 and FG5,} which are synchronised to FG1, independently apply rf signals to the radial rf electrodes and the inner electrode\red{s} of the \red{LEA and} UEA respectively. \red{FG4 and FG5 are out of phase by 180}$\red{^\circ}$. The amplitudes and phases \red{of the function generators} are digitally controlled by a PC. The schemes and results are detailed in the following sub-sections.
	
	\subsection{Matching the phases of the additional rf signals}
	\label{sec:match-phas-addit}
	
\begin{figure}[h]
		\begin{center}
			\subfigure[]{
				{\includegraphics[width=0.54\linewidth]{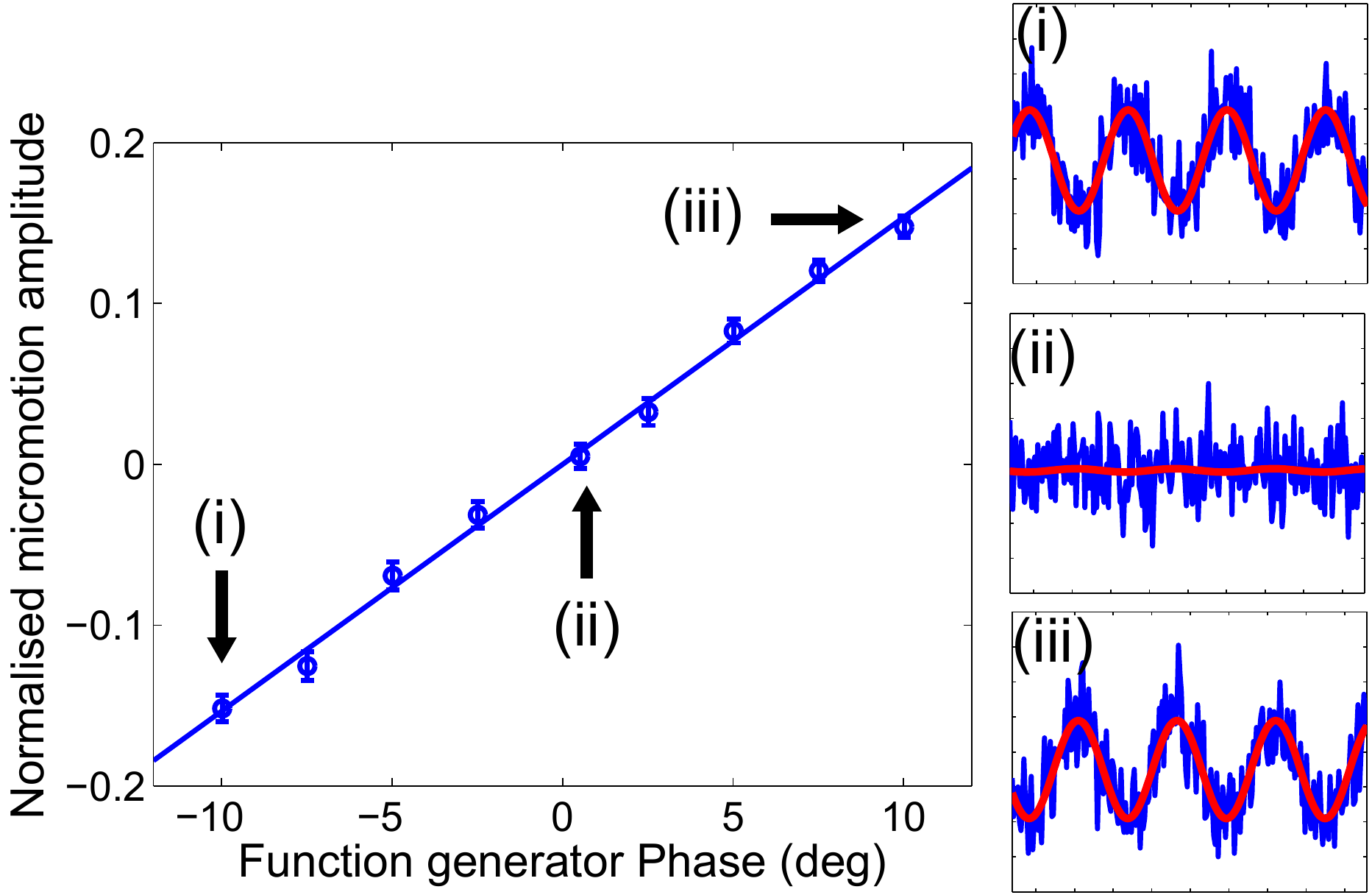}}}
			\subfigure[]{
				\hspace{0.2cm}	{\includegraphics[width=0.4\linewidth]{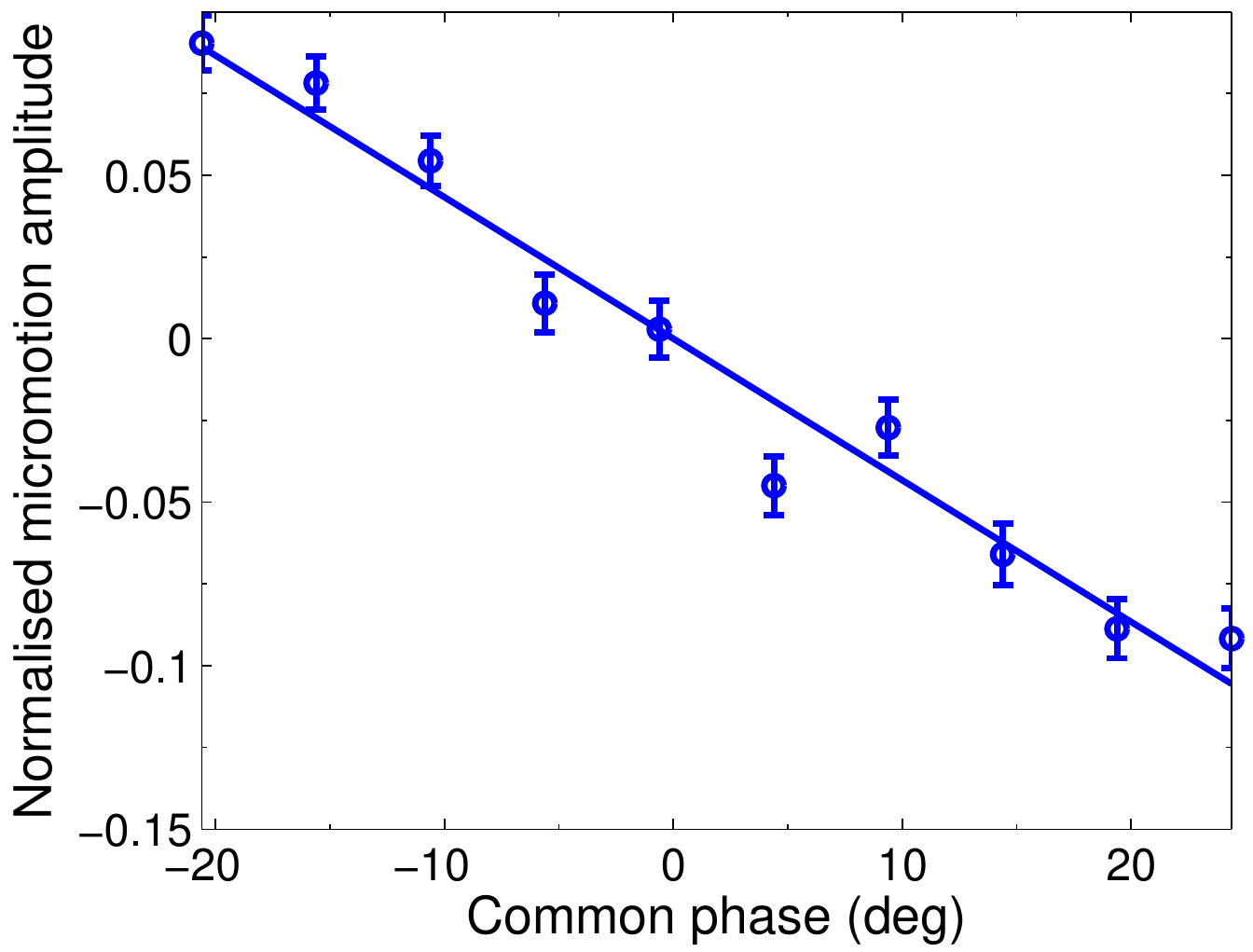}}}
			\caption{(a) Normalised micromotion amplitude as a function of the phase of FG3. The solid line is a linear fit to the data. Also shown in the right column are the correlated fluorescence counts at the individual points (i), (ii) and (iii). The micromotion amplitude is deduced by the sinusoidal fit at the trap frequency as shown in the red lines.  The change of the polarity is clearly visible between (i) and (iii). (b) Normalised micromotion amplitude as a function of the \red{common} phase of FG4 \red{and FG5}, detected with the cavity emission counts at the SPCM. \red{The error bars are the sinusoidal fitting errors of the correlated fluorescence counts.}}
			\label{fig:micromotion-vs-phase}
		\end{center}
	\end{figure}

	The additional rf signals need to be applied in-phase with the main rf drive at the trap. Otherwise they would cause the ion to have excess micromotion due to the phase mismatch \cite{Berkeland:98}. This means that the phase mismatch can be detected through the ion's micromotion, which allows an \textit{in-situ} optimisation of the phases regardless of various phase delays on the transmission lines. Having compensated the micromotion from stray dc fields, the excess micromotion caused by the phase mismatch, $\delta$,  of an additional rf field is described by \cite{Berkeland:98}
	\begin{align}
		\Delta z = -\frac{1}{4}qR\alpha\delta\sin \omega t, 
	\end{align}
	when $\delta \ll 1$ and where $q$ \red{is} the trap's q-parameter, $R$ the ion-electrode distance, $\alpha$ the dipole moment of the trap and $\omega$ the drive frequency used. Therefore the induced micromotion is proportional to the mismatch $\delta$ and notably the polarity of the oscillation changes around $\delta = 0$. This can be clearly seen in \figref{fig:micromotion-vs-phase}(a), where the micromotion amplitude is measured with the rf-correlation technique using the ion's UV fluorescence as the phase of FG3 is scanned. 
	
	The same procedure can be repeated for the other radial and axial directions. In the case of the axial direction, however, we use the ion's cavity emission signal instead of the free-space fluorescence because it has a greater sensitivity to the axial micromotion. In the presence of induced micromotion along the cavity axis, the ion's coupling strength to the cavity mode is modulated at the trap drive frequency \cite{chuah2013detection} and so is the count rate of the cavity emission detected at the SPCM. In order to generate a continuous stream of photon emission, the cavity is locked to the 866 nm transition to satisfy a cavity-assisted Raman resonance condition while the ion is continuously cooled with the 397 nm, 850 nm and 854 nm beams. \red{The ion is placed at the position with the highest ion-cavity coupling gradient to maximise the sensitivity to micromotion.}  \figref{fig:micromotion-vs-phase}(b) shows the micromotion amplitude as a function of the \red{FG4 and FG5 common phase (i.e. the phases of FG4 and FG5 are changed whilst keeping the phase difference constant)}. 
	The results in \figref{fig:micromotion-vs-phase}(a) and (b) show that the phases of the additional rf signals can be unambiguously optimised with the crossing points at the zero amplitude, which ensures minimised micromotion in all three dimensions.
	\subsection{Mapping the cavity mode}
	Using the method discussed in the preceding subsection, the ion can be spatially displaced without inducing excess micromotion by changing the amplitudes of the additional rf signals. Furthermore, the spatial structure of the cavity mode can be probed with the change of the cavity emission counts as the ion is translated along one of the axes. This provides a means to identify the ion's position with respect to the centre of the cavity mode and optimise the coupling between the ion and cavity.      
	
\begin{figure}
		\centering
		\includegraphics[width=0.7\linewidth]{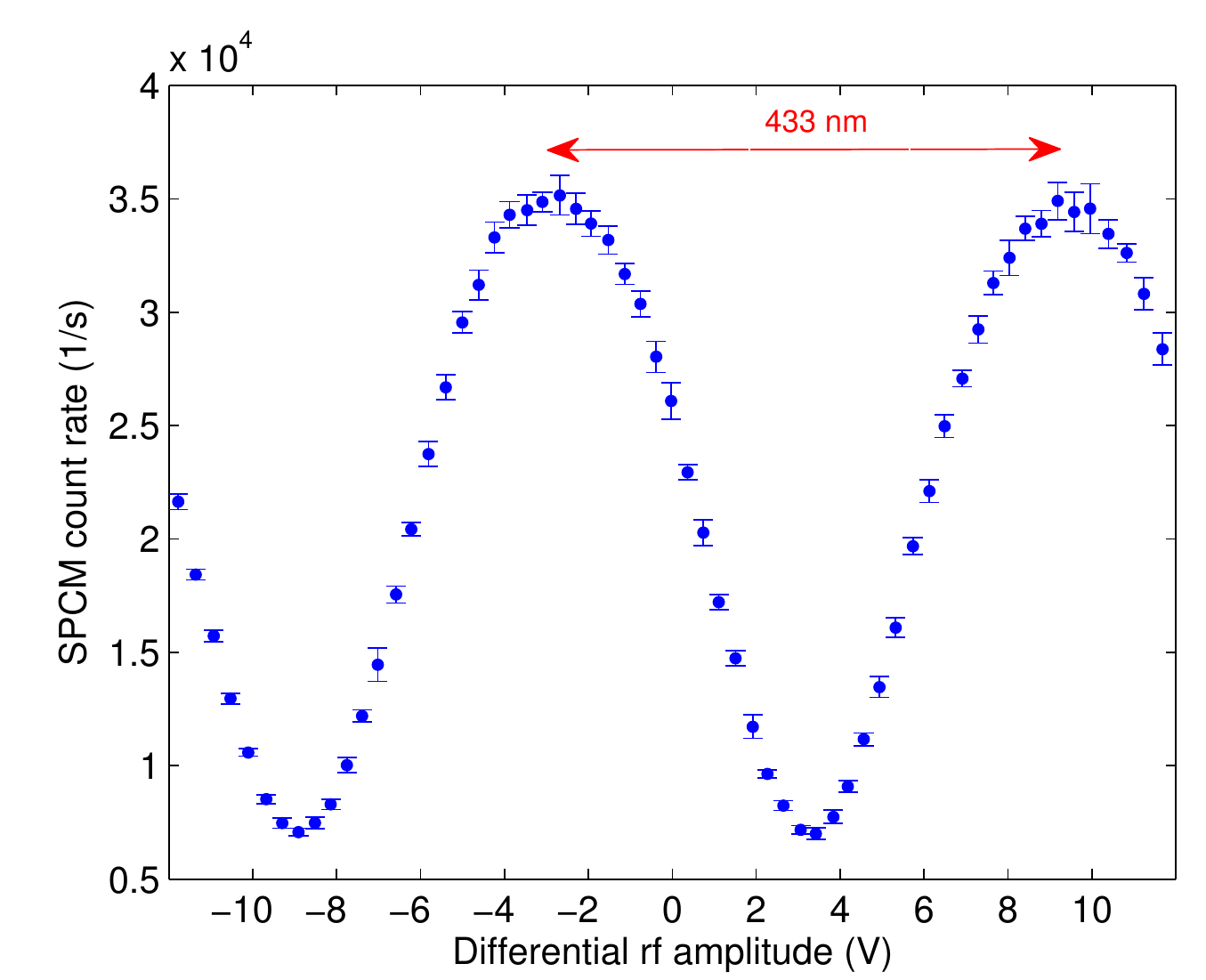}
		\caption{The cavity emission as the ion is being moved along the standing wave of the cavity mode. \red{The x-axis is the differential voltage applied to the upper and lower inner electrodes as probed by the oscilloscope (see Fig.} \ref{fig:MovingIonWithRFSetup_JMO}\red{). Note that this voltage is not of the same magnitude as the actual voltages on the inner electrodes due to inline electrical losses.} The error bars are standard deviations from \red{10} measurements. The separation of the two anti-nodes correspond to a spatial separation of 433 nm as indicated by the red arrow.}
		\label{fig:zrfscan19}
	\end{figure}

	To probe the standing wave structure of the cavity mode, the \red{differential} rf voltage\red{s} of FG4 \red{and FG5}  \red{are} varied to scan the ion's position along the axial direction (see \figref{fig:MovingIonWithRFSetup_JMO}).
	The cavity emission is generated by the cavity-assisted Raman transition as described in the last section and monitored at the SPCM.  
	The result is illustrated in \figref{fig:zrfscan19} which clearly shows a trace of cavity's standing wave, the periodicity of which corresponds to the spatial separation of half the wavelength of 866 nm. There is a background contribution count rate of \red{4,200} counts from the stabilising. We compute a visibility of \red{83}$\red{\pm}$\red{3}\%.
	From the data we can infer that the spatial resolution for the average position of the ion is better than 10 nm.
	
	%CAPTION (ALT)

\begin{figure}[h]
		\centering
		\subfigure[]{
			{\includegraphics[width=0.45\linewidth]{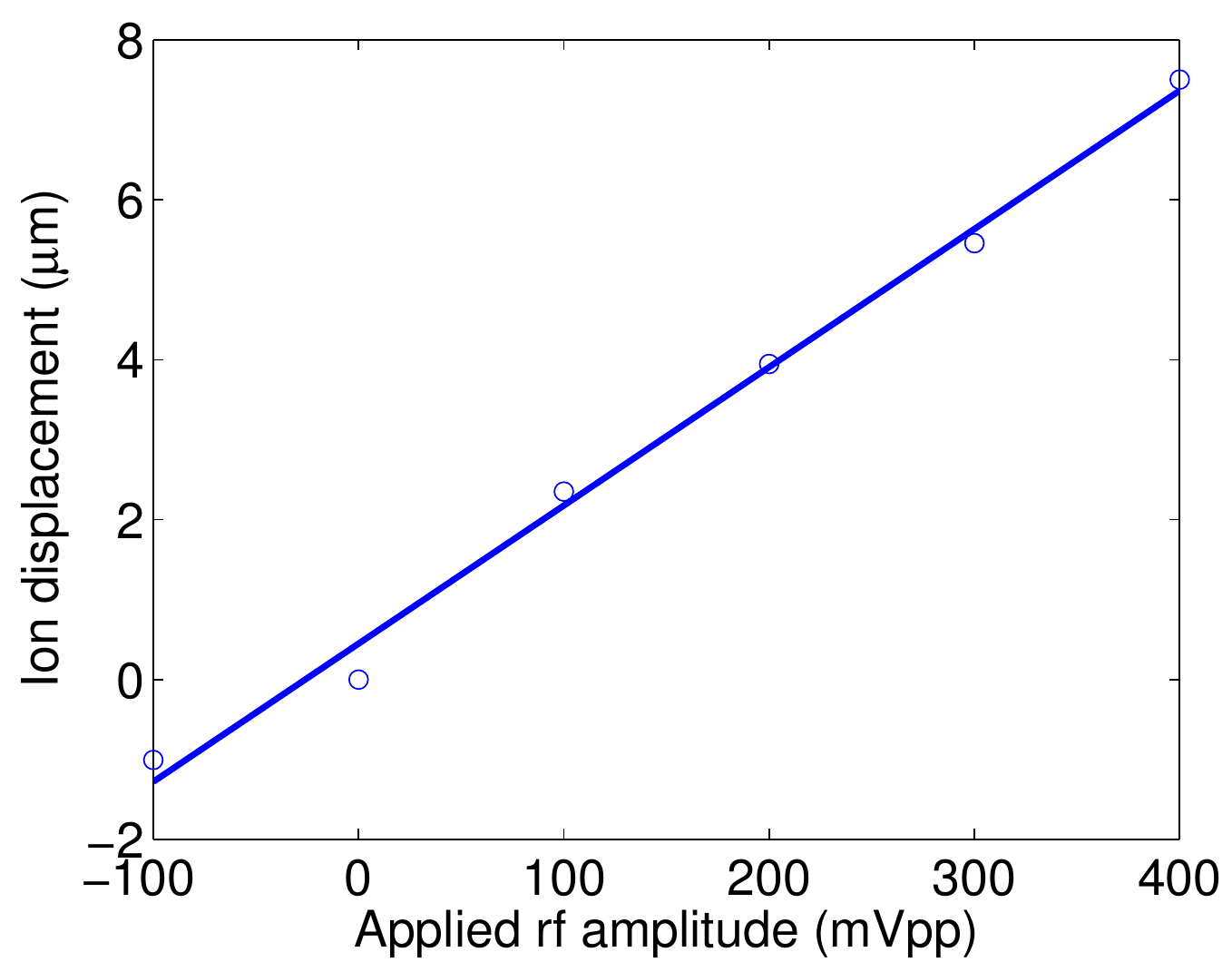}}}
		\subfigure[]{
			{\includegraphics[width=0.45\linewidth]{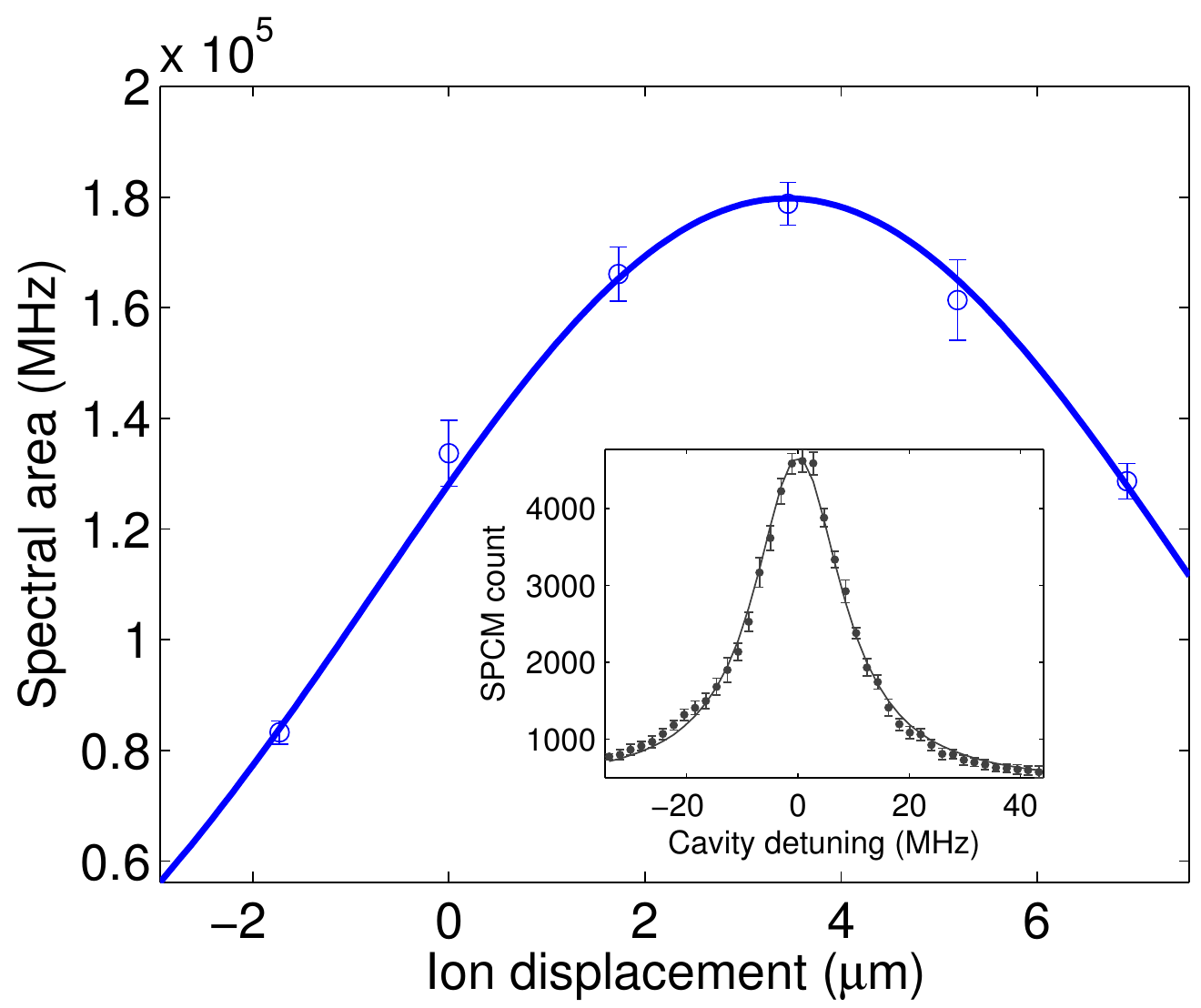}}}
		\caption{(a)The ion's displacement as a function of the rf voltage setting on Function Generator 3 (FG3). The data points are the relative positions of the ion obtained by fitting a Gaussian function to the EMCCD image data. The fitting errors are small for representation here. The line is a fit with a gradient of 17.3 \mum{}/Vpp. (b) The radial cavity mode profile obtained by plotting the spectral area of the cavity length scans at different FG3 amplitude settings. The inset figure shows the cavity emission as the cavity length is scanned across the Raman resonance for an ion displacement of 0 \mum{}. The data is fitted with a Voigt function which is then used compute the spectral area.}
		\label{fig:pos_v_rf_and_radial_profile}
	\end{figure}

	As in the case for the axial direction, the radial profile of the cavity mode can be similarly probed by scanning the amplitude of the rf signal applied to one of the radial electrodes. 
	Due to the relatively large spatial structure of the cavity mode along the radial direction, we can gauge the physical dimension of the profile via the actual displacement of the ion's image on the EMCCD camera (\figref{fig:calcium40_and_optics}). Using the known distance between the UEA and LEA, the camera image is calibrated to 2.0 $\mu$m per pixel.
	Fitting the fluorescence image on the camera with a Gaussian function, the ion's mean position is obtained.  \figref{fig:pos_v_rf_and_radial_profile}(a) shows the ion's position as a function of the rf voltage of FG3 (before the amplifier). For each rf voltage setting, we place the ion at an anti-node of the cavity mode. Subsequently we scan the cavity frequency across the Raman resonance and record the cavity emission spectrum as shown in the inset of \figref{fig:pos_v_rf_and_radial_profile}(b). 
	%Throughout, the ion is continuously cooled and repumped with the 850 nm and 854 nm beams (\figref{fig:calcium40_and_optics}). The 866 nm cavity emission is collected at the SPCM (see inset of \figref{fig:pos_v_rf_and_radial_profile}(b)).
	The spectral area is computed  as an indicator for the relative coupling between the ion and cavity. 
	The spectral area is used since it is immune to potential inhomogeneous broadening and therefore gives more reliable insight  than the SPCM count rate at any given detuning.
	\figref{fig:pos_v_rf_and_radial_profile}(b) shows the spectral area as a function of the ion's radial position. A Gaussian fit gives a waist of $(8.4 \pm 0.1) \mu\mr{m}$ which agrees well with the calculated value of 8.5 \mum{} for the expected waist of the TEM00 mode. When no additional rf is applied, the ion is trapped 3.9 \mum{} from the cavity mode centre in this direction.

	\section{Conclusion}
	
	We have designed and built a novel ion trap with an integrated FFPC.
	By shielding the fibres inside tubular electrodes the detrimental effects on the trapping capabilities by the proximity of the fibres to the trapped ion have been mitigated.

	By integrating additional electrodes into the design, we have succeeded in full control of the ion's position relative to the cavity mode solely by using radiofrequency signals. We have demonstrated that the excess micromotion induced by the phase mismatch of the additional rf signals can be explicitly minimised in all three spatial dimensions.  

	Furthermore, by probing the spatial structure of the cavity mode with the ion's cavity emission, it has been shown that the ion can be positioned with precisions on the order of 10 nm, which enables precise tuning of the ion-cavity coupling.

	This state of the art design offers a promising solution to advancement in the implementation of large scale quantum networks using fibre-linked modular ion traps, an implementation considered  key to advancing experimental quantum information science.
	%Furthermore, this system provides a tunable ion-cavity coupling offering the ability to study atom-light interaction in different coupling regimes.
	Future experiments will look at incorporating FFPCs with mode matching optics \cite{Gulati2017} into this type of ion trap to facilitate highly efficient ion-ion coupling between fibre linked traps.

	\section*{Data Availability}
	
	The data used in this publication can be accessed at
	
	\noindent https://doi.org/10.25377/sussex.5515045
			
	\section*{Acknowledgments}
	We gratefully acknowledge support from EPSRC
	through the UK Quantum Technology Hub: Networked Quantum Information Technologies
	(EP/M013243/1 and EP/J003670/1). E.K. would like to thank late Danny Segal for his wisdom and generosity during the short period he spent under Danny's tutelage.   

	\bibliographystyle{tfp}
	\bibliography{RfTrapPaperBib}

\end{document}